\documentstyle[aps,preprint]{revtex}
\tightenlines
\begin{document}

\title{CLUSTER EMISSION OF $^8$Be IN THE $^{28}$Si+$^{12}$C FUSION REACTION
AT LOW TEMPERATURE.}

\author{M. Rousseau$^1$, C. Beck$^1$, C. Bhattacharya$^{1,5}$, V. Rauch$^1$, S.
Belhabib$^2$, A. Dummer$^3$,R.M. Freeman$^1$, A. Hachem$^2$, R. Nouicer$^1$, D.
Mahboub$^1$, E. Martin$^2$, S.J. Sanders$^3$ and O. Stezowski$^1$, A. Szanto de
Toledo$^4$} 

\address{1) IReS, UMR7500, IN2P3-CNRS-ULP, F-67037 Strasbourg Cedex 02, France}
\address{2) Universi\'e de Nice-Sophia-Antipolis, Nice, France }  
\address{3) University of Kansas, Lawrence, Kansas 66045, USA }  
\address{4) Instituto de fisica da universidade de S\~ao Paulo, S\~ao Paulo,
Brazil} 
\address{5) Variable Energy Cyclotron Centre, 1/AF Bidhan Nagar, Calcutta 700064,India}
%%%%%%%%%%%%%%%%%%%%%%%%%%%%%%%%%%%%%%%%%%%%%%%%%%%%%%%%%%%%%%
% You may repeat \author \address as often as necessary      %
%%%%%%%%%%%%%%%%%%%%%%%%%%%%%%%%%%%%%%%%%%%%%%%%%%%%%%%%%%%%%%

\date{\today}
\maketitle

\newpage

\begin{abstract}
{Inclusive as well as exclusive energy spectra of the light charged
particles emitted in the  $^{28}$Si ($E_{lab}$=112.6 MeV) + $^{12}$C reaction
has been measured using the {\bf ICARE} multidetector array. The data have been
analysed by statistical-model calculations using a spin-dependent level density
parametrization. The results suggest significant deformation effects at high
spin and cluster emission of $^8$Be.} 

\end{abstract}

\vskip 2.0cm

{ PACS numbers: 25.70}

\newpage

\section{INTRODUCTION}

In recent years, extensive efforts have been made to understand the decay of
light di-nuclear systems (A$<$60) formed through low-energy heavy-ion
reactions~\cite{beck}. In most of the reactions studied, the properties of the
observed fully energy damped yields have been successfully explained in terms
of either a fusion-fission (FF) mechanism or a deep-inelastic (DI) orbiting
mechanism behavior. The strong resonance-like structures observed in elastic
and inelastic excitation functions of $^{24}$Mg+$^{24}$Mg  and
$^{28}$Si+$^{28}$Si have indicated the presence of shell stabilized, highly
deformed configurations in the $^{48}$Cr and $^{56}$Ni compound systems
respectively~\cite{beck}. The present work aims to investigate the possible
occurence of highly deformed configurations in the $^{40}$Ca di-nucleus
produced in the $^{28}$Si+$^{12}$C reaction through the study of light charged
particle (LCP) emission. 

\bigskip

\section{EXPERIMENTAL PROCEDURES}

The experiment was performed at the IReS Strasbourg VIVITRON tandem facility
using 112.6 MeV $^{28}$Si beams on a $^{12}$C(160 $\mu$g/cm${^2}$) target. Both
the heavy ions and their associated LCP's were detected using the {\bf ICARE}
charged particle multidetector array~\cite{bello}. The heavy ions were
detected in eight telescopes, each consisting of an ionisation chamber  (IC)
followed by a 500 $\mu$m Si detector. The in-plane coincident LCP's were
detected using four triple telescopes ( Si 40 $\mu$m, Si 300 $\mu$m, 2 cm
CsI(Tl)), 16 double telescopes ( Si 40 $\mu$m, 2 cm CsI(Tl)) and two double
telescopes (IC, Si 500 $\mu$m) located at the most backward angles. Typical
inclusive $\alpha$ energy spectra are shown in Fig.1.a. 

\newpage

\section{EXPERIMENTAL RESULTS AND STATISTICAL-MODEL CALCULATIONS}

The data analysis was performed using CACARIZO~\cite{viesti}, the Monte Carlo
version of the statistical-model code CASCADE. The angular momenta
distribution, needed as the main input to constrain the calculation, was taken
from $^{28}$Si+$^{12}$C complete fusion data~\cite{harm,vien}. The other
ingredients such as the nuclear level densities and the barrier transmission
coefficients, are usually deduced from the study of the evaporated LCP spectra.
Standard statistical-model calculations are not able to reproduce the shape of
experimental $\alpha$-particle energy spectra satisfactorily~\cite{viesti}.
Several attempts have been made to explain this anomaly either by changing the
emission barrier or by using a spin-dependent level density. In hot rotating
nuclei forme density at higher angular momentum should be spin dependent. In
CACARIZO, the level density, $\rho(E,J)$, for a given angular momentum $J$ and
energy $E$ is given by the well known Fermi gas expression: 
$$\rho(E,J)=\frac{(2J+1)}{12}a^{1/2}(\frac{\hbar^2}{2{\Im}_{eff}})^{3/2}\frac{
1}{(E-\Delta-E_J)^2}exp(2[a(E-\Delta-E_J)]^{1/2}),$$ 
Where $a$ is the level density parameter, $\Delta$ is the pairing correction,
$E_J=\frac{\hbar^2}{2{\Im}_{eff}}J(J+1)$ and ${\Im}_{eff}={\Im}_{0}\times
(1+\delta_1J^2+\delta_2J^4)$ with ${\Im}_{0}$ the rigid body moment of
inertia and $\delta_1, \delta_2$ the deformability parameters. By changing
the deformability parameters one can simulate the deformation effects on the
level densities.  For the $^{28}$Si + $^{28}$Si reaction~\cite{chan}, the shape
of the inclusive and exclusive $\alpha$ energy spectra are well reproduced
by using large deformation effects~\cite{chan}. Similarly the
experimental inclusive $\alpha$ energy spectra for $^{28}$Si + $^{12}$C of 
Fig.1.a are better described by using deformation effects (dotted lines) than
with the standard liquid-drop deformation (dashed lines).\\ 

The exclusive energy spectra of $\alpha$-particle in coincidence with
individual S and P ER's shown in Figs.1.b and 1.c are quite interesting. The
dotted lines are the predictions of CACARIZO using non-zero values of the
deformability parameters. The energy spectra associated with S are completely
different from those associated with P. The latter are reasonably well
reproduced by the CACARIZO curves whereas the model could not predict the shape
of the spectra obtained in coincidence with S (Fig.1.b). This is due to the
fact that an additional component might be significant in this case. One could
suggest the hypothesis of a contribution arising from the decay of unbound
$^{8}$Be produced in a binary reaction $^{40}$Ca $\rightarrow$
$^{32}$S+$^{8}$Be. In order to determine the sources of both the $\alpha$
emission and $^{8}$Be breakup the invariant cross sections in coincidence with
P and S are plotted in Fig.2 in the (V$_{\perp}$,V$_{\parallel}$) plane.
Fig.2.b shows the invariant cross sections in coincidence with P which maxima
are centered on the compound nucleus velocity as expected for a
fusion-evaporation mechanism. Fig.2.a presents two additional contributions
(in circles) for angles close to 30$^{\circ}$ and 60$^{\circ}$ arising from the
binary decay of unbound $^8$Be. This conclusion is also consistent with the ER
kinematical analysis of the S and P exclusive energy spectra (not shown here).
The question of the real nature (FF or orbiting) of this decay process remains
to be explored. 

\section{SUMMARY}

The $\alpha$-particle energy spectra measured in coincidence with S have an
additional component which may come from the decay of $^8$Be, which is unbound
and produced through the binary decay of $^{40}$Ca $\rightarrow ^{32}$S +
$^8$Be. Work is in progress to analyse the proton energy spectra as well as the
in-plane angular correlations of both the $\alpha$-particles and the protons.

\begin{figure}
\caption{Inclusive (1.a) and exclusive experimental (solid line) $\alpha$
energy spectra in coincidence with S (1.b) and P (1.c). The dotted and
dashed lines are CACARIZO calculations with and without deformation.}
\label{fig.1} 
\end{figure}

\begin{figure}
\caption{$\alpha$ invariant cross section in the (V$_{\perp}$,V$_{\parallel}$)
plane in coincidence with S (2.a) and P (2.b), the two circles in (2.a) show 
the additionnal contribution arising from the decay of $^8$Be.}
\label{fig.2}
\end{figure}


\begin{references}
\bibitem{beck} S.J. Sanders, A. Szanto de Toledo, and C. Beck, {\it Phys. Rep.}
{\bf 311}, 487 (1999). 
\bibitem{bello}T. Bellot, Ph.D. Thesis, Strasbourg University, Report {\bf
IReS 97-34}. 
\bibitem{viesti}G. Viesti, B. Fornal, D. Fabris, K. Hagel, J.B. Natowitz,
G. Nebbia, G. Prete, and F. Trotti, Phys. Rev. C {\bf 38}, 2640 (1988).
\bibitem{harm}B.A. Harmon, B.A. Harmon, S.T. Thornton, D. Shapira, J.
Gomez del Campo, M. Beckerman, {\it Phys. Rev.} C {\bf 34}, 552 (1986).
\bibitem{vien}M.F. Vineyard, J.F. Mateja, C. Beck, S.E. Atencio, L.C. Dennis,
A.D. Frawley, D.J. Henderson, R.V.J. Janssens, K.W. Kemper, D.G. Kovar, C.F.
Maguire, S.J. Padalino, F.W. Prosser, G.S.F. Stephans, M.A. Tiede, B.D.
Wilkins, and R.A. Zingarelli, 
{\it Phys. Rev.} C {\bf 47}, 2374 (1993).
\bibitem{chan}C. Bhattacharya, M. Rousseau, C. Beck, V. Rauch, S. Belhabib, A.
Dummer, R.M. Freeman, A. Hachem, R. Nouicer, D. Mahboub, E. Martin, S.J.
Sanders, O. Stezowski, and A. Szanto de Toledo, 
{\it Nucl. Phys.} {\bf A}, (1999) {\it in press}.

\end{references}
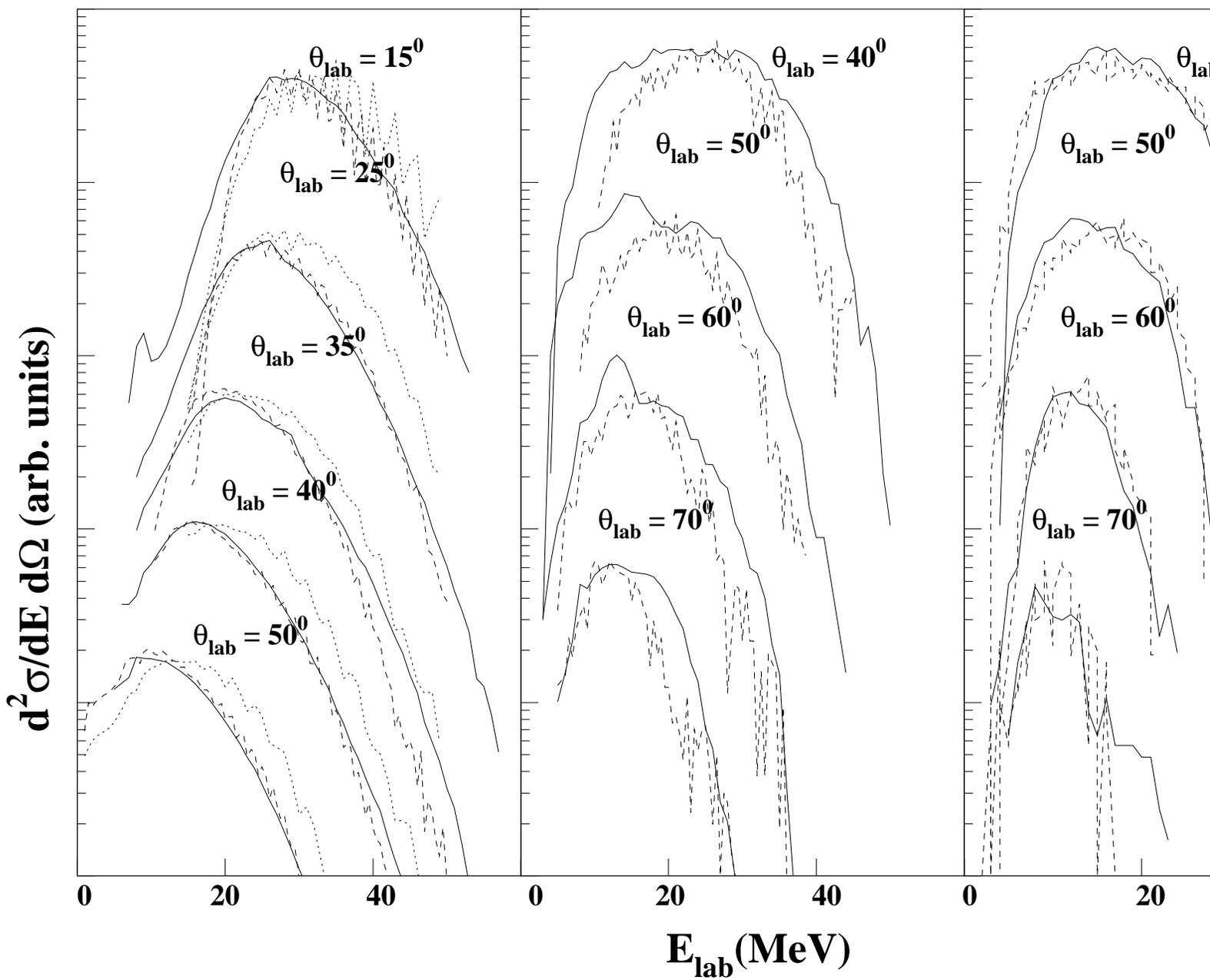
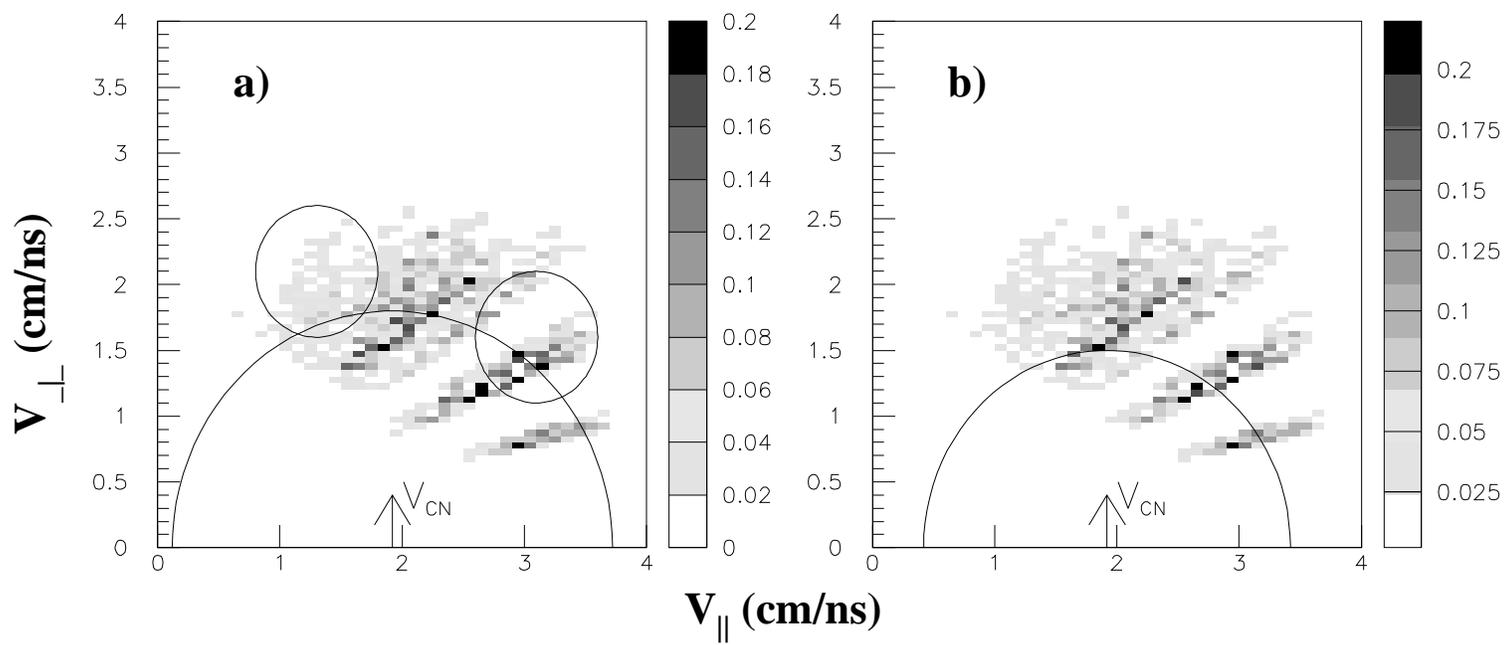
\end{document}